\documentclass[aps,floatfix,nofootinbib,preprint]{revtex4}
\usepackage{graphicx}
\usepackage{amssymb}
\usepackage{amsmath}
\usepackage{bbold}
\usepackage{amsfonts}
\usepackage{amsbsy}
\usepackage{color}
\def\lb{\langle}
\def\rb{\rangle}
\def\be{\begin{equation}}
\def\ee{\end{equation}}

\def\half{{\textstyle\frac{1}{2}}}

\begin{document}

\title{Force and pressure in many-particle quantum dynamics}
\author{G.F.~Bertsch}

\affiliation{Department of Physics and Institute for Nuclear Theory,     
University of Washington, Seattle, Washington 98915, USA  
}  

\begin{abstract}

The Newtonian concept of force may be useful in some aspects of
the dynamics of many-particle quantum systems such as fissioning
nuclei. Following Ehrenfest's method, we show that the quantum
kinetic force between parts of an extended quantum system can be
described by an operator acting on the boundary between the two
subsystems.  The contribution to the force due to a short-ranged
particle interaction can also be treated in the same way.
This includes interaction effects treated in
density functional theory by local functionals.  The force
operators are applied to several simple models to demonstrate
the method.

\end{abstract}
\maketitle 
 
\section{Introduction}
Ehrenfest's celebrated theorem showed that the classical Newton
concept of force applies equally well to the acceleration of 
isolated quantum systems, but what about forces between parts
of extended quantum systems?  The question is relevant to 
the description of the dynamics of nuclear fission. At some
point the nucleus is elongated into
a shape resembling the nascent fission fragments joined by
a small neck.  The Coulomb force between the two nascent
fragments is counterbalanced by an attractive nuclear force
transmitted through the neck region; scission only occurs
when neck force is sufficiently weakened.

In this work, we show that in an extended system the quantum force 
of one part on the rest can be consistently defined.
In general, the interaction has long-range components
such as the Coulomb and short-range ones such as the exchange-correlation
energy in density-functional theory.  Except for the
long-range components of the interaction, the  force
reduces to an  operator acting at
the interface between the two subsystems. The operator
contains derivatives of the wave function at the interface
due to the kinetic term in the Hamiltonian.

We first consider the Schr\"odinger equation for a  particle in one 
dimension.  We then generalize the operator scope  to many-particle
systems in three dimensions.

\section{Particle in one dimension}
\subsection{Kinetic force}
Consider a particle in one dimension and governed by the Schr\"odinger
Hamiltonian $H = K + V$. Here $K$ is the kinetic Hamiltonian
\be
K = -{\hbar^2\over 2M} {\partial^2\over\partial x^2}.
\label{K-def}
\ee  
where $M$ is the mass of the particle.
$V$ in the Hamiltonian is a potential depending on $x$. We divide 
the space
into two subsystems at some point $x_0$.  The particle number
$n_{R,L}$ on each side and the positions of their centers of
mass $x_{cm(R,L)}$ are evaluated by restricting the integrations
over the wave function 
to one side or
the other of the point $x_0$.  For the right hand side, this is
achieved by the operators

\be
\hat {\mathbb{1}}_{R} = \Theta(x-x_0)
\ee
\be
\hat x_R = x \Theta(x - x_0)
\ee
where $\Theta$ is the Heaviside step function.  We write the expectation values
in the wave function $\phi(x)$ as 
\be
\langle \phi | \hat {\mathbb{1}}_R | \phi\rangle = n_R
\ee
\be
\label{xR}
\langle \phi | \hat x_R | \phi \rangle = n_R x_{cm(R)}.
\ee

Depending on the wave function, both of these quantities can
vary with time under the Schr\"odinger dynamics.  However, it is
not useful to call the effect on a subsystem a force if the
particle number is changing. We therefore restrict the
definition to wave functions for which $ d n_R / d t$ and $d^2
n_R / d t^2$ are zero at the time when the force is computed. 
Then the time dependence of Eq. (\ref{xR}) will provide the
acceleration $a_R$ of the right-hand side computed as
\be
a_R = {1\over n_R} {d^2 \over d t^2} \langle \phi | \hat x_R | \phi
\rangle.  
\ee
Given the acceleration, he Newtonian force can be defined as 
\be
\label{F-def}
F_R = M {d^2 \over d t^2} \langle \phi | \hat x_R | \phi \rangle. 
\ee
We can now carry the
derivatives in the Heisenberg representation by twice applying the
usual commutator formula
\be
{d \over dt} \langle \hat {\cal O}\rangle   = {1\over i\hbar} \langle 
[{\cal O},H]\rangle.
\ee
The resulting operator for the kinetic energy term in $H$  is
\be
\label{FK-op}
\hat F_k(x_0)  =  -M[K,[K,\hat x_R]] =
 -{\hbar^2 \over 4 M}
\left(2{\overleftarrow\partial \over \partial
x}\delta(x-x_0){\overrightarrow\partial \over \partial x} 
- {\overleftarrow {\partial^2}  \over \partial x^2}\delta(x-x_0)
-\delta(x-x_0){\overrightarrow {\partial^2}  \over 
\partial x^2}\right)
\ee
	
To show how Eq. (\ref{FK-op}) works, we apply it to some very
simple Hamiltonians.  The first is a particle in a box.  The
n-th stationary state in the box has the wave function
\be
\phi = \sqrt{2\over L} \sin (k_n x )
\ee
where $L$ is the length of the box and $k_n = n\pi/L$.  Its
energy is
\be
E_n = {\hbar^2 k_n^2 \over 2 M}.
\ee
The force exerted by the particle on the boundary can be 
calculated by the usual relation between force, energy,
and displacement
\be
\label{F-box}
F = - {d E_n(L)\over d L} = {\hbar^2 \pi^2 n^2\over M L^3}.
\ee
This force must be sustained throughout the interior of the box.  If we
arbitrarily divide the space in two the same force must act between
the two sides.  Applying Eq. (\ref{FK-op}) we find
\be
F_R = {\hbar^2\over M L} \left(\left( \left. {d\sin (k_n x)\over dx}
\right|_{x_0}\right)^2
 - \sin(x_0) \left. {d^2 \sin(k_n x)\over d x^2}\right|_{x_0}\right)
=  {\hbar^2 \pi^2 n^2\over M L^3}
\ee
in agreement with Eq. (\ref{F-box}).

Another simple example is the expansion of a Gaussian wave packet.
The initial wave function may be written
\be
\label{phi-ho}
\phi(x) = {1\over b^{1/2}\pi^{1/4}} e^{- x^2/2 b^2}.
\ee
Here $b$ is a parameter controlling the width of 
the Gaussian.  The time-dependent Schr\"odinger equation has the
exact solution
\be
\label{phit}
\phi(x,t) = {1\over b^{1/2}\pi^{1/4}}(1 + i \hbar t/M b^2)^{-1/2} \exp(
-{\textstyle\frac{1}{2}}  x^2/
 b^2(1+i \nu t/M))).
\ee
There is no particle transfer at $x_0=0$ so we can calculate the
force at that point. The center-of-mass position of the right-hand
side particle density is
\be
\label{xRt}
\lb x_{R}\rb = \int_{x_0}^\infty dx x |\phi(x,t)|^2 = {b \over 2 }
\pi^{1/2} (1 + (\hbar t/Mb^2)^2)^{1/2} 
\ee
By explicit differentiation of Eq. (\ref{xRt}), the acceleration of
right-hand density distribution is
\be
\label{ddotxR}
{d^2 \over d t^2}\lb \hat x_R \rb = {1 \over 2 \pi^{1/2} M^2b^3} (1 + (\hbar
t/M b^2)^2)^{-3/2}.
\ee
The force calculated from this acceleration together with  Eq. (\ref{F-def})
agrees with that obtained by the expectation value of 
Eq. (\ref{FK-op}) in the wave function $\phi(x,0)$.
The important point is that Eq. (\ref{FK-op}) only 
requires information about the
wave function around the point $x_0=0$, and yet it 
perfectly describes the cm acceleration
of all the matter to the right of that point.

\subsubsection{Potential contribution}
Now we add an external potential field $V$ to the Hamiltonian and treat the
associated force $F_V$ acting on a region in the same way.  
The operator requires the double commutator
\be
\hat a_R = -\frac{1}{\hbar^2} [V,[K,\hat x_R]]
\ee
and is evaluated as
\be
\label{FV-def}
\lb \hat a_R\rb =\int_{x_0}^\infty dx\, \rho(x) {\partial V \over \partial
x};
\ee
Not surprisingly, this formula
is very similar to 
Ehrenfest's second equation, $d \langle p \rangle /dt = 
-\langle d V/dx \rangle$. To illustrate 
the application of
Eq. (\ref{FV-def}), consider a particle in the ground state of the
harmonic oscillator Hamiltonian.  It is convenient to express the
potential in the form
\be
V(x) = \frac{\hbar^2}{2M b^4}   x^2. 
\ee
Its ground state is the wave function of Eq. (\ref{phi-ho}).
Since it is stationary, the acceleration is zero everywhere and there
should be an exact force balance
at all points:
\be
F_K + F_V = 0 \,\,\,\,\,\  {\rm for~ all} \,\,x_0.
\ee
It is an easy exercise to verify that 
\be
\label{FK-ho}
F_K =   {\hbar^2 \nu^{3/2} \over 2 M \pi^{1/2}} e^{-\frac{1}{2}\nu x_0^2}.
\ee
Carrying out the integration
in Eq. (\ref{FV-def}),  one obtains the negative of Eq. 
(\ref{FK-ho}).  This  verifies the
force balance for the ground state, but in fact
it must be true for excited states as well.

\subsection{Beyond the one-particle Hamiltonian}
The formulas of the last subsections are  easily generalized to
multiparticle systems when particles interact through  mean-field
potentials.  For the kinetic quantum force, all the physics is governed
by one-body operators and so all of the forces are additive.  The summation
can be carried out at the level of the wave function to obtain
the the single-particle density matrix $\rho(x,x')$.  The corresponding kinetic force is then
given by the compact expression
\be
\label{density-matrix}
F_K = -{\hbar^2\over 2 M} \left.{ d^2 \over d s^2} \rho(x_0 +s/2,x_0-s/2)\right|_{s=0}.
\ee
An even more compact expression makes use of the Wigner representation
of the density matrix,
$f(x,p) = \frac{1}{2 \pi}\int_\infty^\infty d s 
\exp(i p s) \rho(x +s/2,x-s/2)$.  Then Eq. (\ref{density-matrix})
becomes
\be
F_K = {\hbar^2\over M}\int_\infty^\infty d p p^2 f(x_0,p)
\ee

To introduce particle-particle interactions, 
we first note that the instantaneous force can be derived by
expanding
the Hamiltonian evolution operator to second order in time. In
mean-field theory the single-particle potential can be expanded
as well. However, the time-dependent corrections to the
instantaneous potential start with terms beyond second order in time
and so can be dropped.
All that is required to treat interaction effects is determine 
their contribution with an instantaneous single-particle potential.

The simplest case to deal with is a two-body finite-range interaction
$v(x -x')$.  By Eq. (\ref{FV-def}) the force $F_V$ may be expressed
\be
F_V = M\int_{x_0}^\infty dx \,\rho(x) \int_{-\infty}^\infty dx'
\rho(x')
{d \over d x} v(x - x').
\ee
Next split the integral over $x'$ into two at the integration point $x$.
The contribution with $x'$ in the range $x<x'<\infty)$ can be shown to 
vanish;
physically the interactions between particles in
the same region do not affect the center-of-mass motion in that region.
Thus $F_V$ reduces to
\be
\label{long-range}
F_V = M \int_{x_0}^\infty dx \,\rho(x) \int_{-\infty}^{x} dx' \rho(x')
{d \over d x'} v(x'' - x') 
\ee
This form is appropriate as it stands for long-range forces such as
the Coulomb interaction.   It needs the entire density distribution
to calculate it, but at least it doesn't require numerical differentiation
of global energies.  

As the range of the interaction decreases, it is clear that only the
the density near $x_0$ contributes to the integral.  An easy way to
derive the force is to follow the spirit of 
density function theory, where the interaction energy density
${\cal V}$ is treated as a local function of position.  
We start with a simple form for the interaction energy functional,
${\cal V} = \frac{1}{2} v_0 \rho^2(x)$.
The corresponding single-particle potential is
\be
V(x) = {d \,{\cal V}(x) \over d \rho} = v_0 \rho(x).
\ee
Inserting this in Eq (\ref{long-range}), we have
\be
\label{FV-v0}
F_V = v_0\int_{x_0}^\infty  dx  \rho(x)  {d\rho \over dx} = 
\half v_0  \int_{x_0}^\infty  dx  {d \over dx} \rho^2(x)  
= -\half v_0  \rho^2(x_0). 
\ee
To obtain the far right-hand equality, 
we have assumed that $\rho \rightarrow 0$ at large
$x$.

Eqs.  (\ref{density-matrix}) and (\ref{FV-v0}) can be easily tested in the
one-dimensional Fermi gas model.   We first derive the force from the 
total energy in a box. The total force on the box wall can
be calculated as before by taking the derivative of the total energy
with respect to box size.  The result is 
\be
E =  {\hbar^2 k_F^2 \over 6 M}N + {v_0 \over 2 L} N^2  
\ee
where $N$ is the number of particles, $k_F \approx  \pi N/L$ and 
$\rho = N/L$.  
 Taking the derivative, the force on the wall is
\be
F_b = -{d E \over d L} =  {\hbar^2 k_F^2 \over 3 M } {N\over L} + \half v_0
{N^2 \over L^2}.  
\ee
The first term is the kinetic contribution;  it may be calculated
from Eq. (\ref{FK-op}) taking $(\rho(x+s/2,x-s/2)= \sin(k_F s)/(\pi s)$.
The second is identical to Eq. (\ref{FV-v0}).

The same method can be used to derive the force associated with
any energy density functional $\cal V$
that can be expanded in powers of the local density $\rho(x)$.
We write 
\be
{\cal V}(\rho) = \sum_m {1\over m!} v_m \rho^m.
\ee
Then $F_V$ can be evaluated similarly to Eq. (\ref{FV-v0}) as
\be
F_V = \sum_m {1\over m!} (m-1) v_m \rho(x_0)^m.
\ee
This can also be expressed as \cite{be77}
\be
F_V = V(x_0)\rho(x_0) - {\cal V}(x_0).
\ee
In the last form, the force is seen to depend only on the density
at the interface.

\section{Three dimensions}

The generalization to three dimensions is trivial if the interface
between the two subsystems is a plane.  Then the kinetic
force operator acts perpendicular to the plane and just requires an
integration over the transverse coordinates.  Formally, one can
define a stress tensor $\Pi$ that transmits momentum from one part
of the system to another.  In the co-moving frame of the medium, 
the stress tensor $\Pi$ associated with density-functional dynamics
is given by an expression very similar to the one-dimensional formula,
\be
\Pi_{ij}(\vec r)  =  {\hbar^2\over M}\nabla_{s_i} \nabla_{s_j} \rho(\vec r - \vec s/2,
\vec r + \vec s/2) + \left(V(\vec r) \rho(\vec r) -{\cal V}(\vec
r)\right)\delta_{ij}.
\ee
The two terms represent the kinetic and interaction contributions,
respectively.
The interaction term is isotropic, but that need not be the case for
the kinetic term.  A perpendicular Newtonian force can be calculated across any
plane by the integral
\be
\vec F\cdot \vec u  = \int \Pi \cdot d \vec A
\label{3D-force}
\ee
where $\vec u$ is a unit vector perpendicular to the plane $\vec A$.
In practice, one would choose a plane going through
the neck region that joins the two nascent fragments.  Note that
there is no mechanism here to generate a transverse force between
subsystems.   

In practice in calculating dynamics in nuclear physics, 
condensed matter physics and 
quantum chemistry, one defines the configurations 
by minimizing an energy functional in the presence of
a fixed external field.
In such situations the wave function has no currents
so the conditions for calculating the
force across a plane are satisfied.  However, the constrained
minimization procedure also permits a third way to calculate
force.  This is to use the Feynman-Hellman theorem and  calculate
the energy derivative as an expectation value of the derivative
of the constraining field.  This is a much easier task than to
explicitly calculate total energy derivatives numerically.  Still,
the quantum force operator might still be useful in some situations
and as an independent check on computations carried out by other
methods.

\section{Appendix}

\subsection{Pairing}

Pairing is very important in low-frequency nuclear dynamics.  
At present, the
most well-justified models incorporating pairing are based on the 
Hartree-Fock-Bogoliubov (HFB) extension of mean-field theory.  It is now possible to carry out the
integration of the HFB equations of motion without uncontrolled 
approximation \cite{bu16}.  It is also clear from simplified implementations
that lifetimes are strongly dependent on the pairing field
\cite{ne78,ro14}.  The force associated with pairing can be derived in the
same way as we treated the ordinary interactions in the Hamiltonian.
The dynamics is governed by the HFB equation of motion, also known as
the Bogoliubov-de Gennes equation.  This is written
\be
i \hbar {d \over d t} \left(\begin{array}{c}
\vec v_\alpha \\
\vec u_\alpha \\
\end{array}
\right)=
\left(\begin{array}{cc}
H & \Delta\\
-\Delta^* &  -H\\ 
\end{array}\right)
\left(\begin{array}{c}
\vec v_{\alpha} \\
\vec u_{\alpha} \\
\end{array}
\right),
\ee
in the usual notation.  
  The pairing energy associated with a two-body
interaction $\bar v$ 
can be expressed $ E_p  = \frac{1}{4} Tr_2 \kappa^* \bar v \kappa $ 
 \cite{RS} where 
$\kappa = \sum_\alpha v_\alpha u^T_\alpha$ ; the 
$\Delta$ field is given by $\Delta = \frac{1}{2} \bar v \kappa$.  The expressions for
the energy and field are very 
similar to those for the ordinary interaction with the replacement of
$\rho$ by $\kappa$.  We expect that the derived force will come out
in a similar way if the interaction $\bar v$ is short-ranged.  Then
the force would be equal to the pairing energy density at the 
division point. For most physical systems, the pairing energy density is 
small compared to  other interaction terms,
so the
pairing force can be neglected in practice.

\subsubsection{Adiabaticity}

For the complete dynamics of 3-D media 
it is essential to understand the time scale of the collective
motion with respect to the time scale to establish local 
equilibrium.  For slow collective motion, the presence of 
interactions beyond mean field keeps the Fermi surface nearly spherical
and the resulting stress tensor is nearly isotropic.  In the opposite limit,
the stress tensor remembers the strain history of the system,
and the Fermi surface can have quadrupolar distortion.
In terms of the Lam\'e parameterization of the stress tensor,
in the adiabatic case the compressibility is governed by a pressure
field
\be
P = \lambda + \textstyle\frac{2}{3} \mu
\ee
whereas in the diabatic case the longitudinal stress for a
strain field in the $z$ direction is given by
\be
\Pi_{zz} = \lambda + 2 \mu.
\ee
The kinetic stress tensor for a diabatically deformed Fermi surface 
has Lam\'e coefficients
\be
\lambda = \mu = {k_f^2 \over 5 M}. 
\ee
A number of Fermionic systems exhibit diabatic
dynamics in high-frequency oscillations.  We mention zero
sound in liquid $^3$He \cite{wi67}, the wave-length dependence
of plasmons in conductors \cite{ja67}, and the giant quadrupole resonance
in large nuclei \cite{lh98}.     

These considerations only indirectly affect the forces we have 
calculated here.  As stated earlier, the force or stress tensor
depends only on the instantaneous state of the system.  If that
wave function is obtained by a constrained mean field solution with 
density constraint operators, it will have no local
currents although  it may have a deformed Fermi surface as a result
of the constraints.  If one releases the constraints and allows
the system to evolve, the proto-fragments will at first be
accelerated away from each other by the force dynamics treated
here.  But later as the state of system changes, the compressibility
will play a role.  A stronger restoring force will be present
in the diabatic dynamics than in the adiabatic.  The considerations
discussed here cannot tell us whether the system eventually
come apart in fragments.

\section{Acknowledgment}  
The author would like to thank T. Kawano, W. Younes, and A.
Bulgac for discussions on nuclear fission dynamics.

\end{document}